\documentclass[amsmath,amssymb,numbers]{revtex4}
\begin{document}

\title{Maximal Entanglement, Collective Coordinates and Tracking the King}

\author{M. Revzen}
\affiliation {Department of Physics, Technion - Israel Institute of Technology, Haifa
32000, Israel}

\date{\today}

\begin{abstract}
Maximal entangled states (MES) provide a basis to two d-dimensional particles Hilbert
space, d=prime $\ne 2$. The MES forming this basis are product states in the collective,
center of mass and relative, coordinates. These states are associated (underpinned) with
lines of finite geometry whose constituent points are associated with product states
carrying Mutual Unbiased Bases (MUB) labels. This representation is shown to be convenient
for the study of the Mean King Problem and a variant thereof, termed Tracking the King
which proves to be a novel quantum communication
channel.\\
The main topics, notions used are reviewed in an attempt to have the paper self
contained.\\

\end{abstract}

\pacs{03.65.Ta;03.65.Wj;02.10.Ox}

\maketitle

\section {  Introduction}

The Mean King Problem (MKP) introduced in \cite{lev} and studied extensively since \cite{durt}, is a
fundamental quantum mechanical problem \cite{werner} in which Alice determines the outcome of the King's
measurement of one particle in an orthonormal basis (b) of his choice subject to the following protocol:
Alice prepares a state of her choosing that the King measures, and is allowed one  control measurement
(that, like the state she prepares, may involve two particles,
one inaccessible to the King) subsequent to the King's measurement. After she completes her control
measurement the King informs her the basis (b) he used  and her task is to specify the
outcome (m) of his measurement. In  Tracking the King, introduced here, her task is to
determine, via her control measurement, the basis used by the King .\\

The analysis we adopt involves two recent studies, both relates to the notions of maximally
(or completely) entangled states (MES) as analyzed, e.g. by \cite{fivel}, and to mutually
unbiased bases (MUB), see, e.g. \cite{tal,wootters4,durt}: The first relates the MES to
collective, viz center of mass and relative coordinates of the constituent (two) particles.
Thus the two particle system may always be accounted for by center of mass and relative
coordinates. We identify product states in the collective coordinates that form MES
\cite{rev1} in the particles coordinates and show that these states provide a MES
orthogonal basis spanning the two d-dimensional particles Hilbert space. This observation
simplifies the analysis considerably. For the sake of completeness and notational clarity
we give in the next section, Section II, a brief review of both the single particle MUB
\cite{tal,durt} and two particles mutual unbiased collective bases, (MUCB), \cite{rev1}.
The second study involves the association of MES with finite geometry \cite{rev2}. This
allows an intuitive interpretation to the solution of the original Mean King Problem (MKP)
\cite{lev,berge,durt} and visualization of the present Tracking Problem. In essence, the
geometrical approach associates (underpins) states or operators in Hilbert space with lines
and points of the geometry . e.g. the approach was applied to {\it one} particle
d-dimensional Hilbert space and provided a convenient formulation of finite dimensional
Radon transformation \cite{rev2}. Here we extends this single particle study to two
particles systems. In so doing we identify what is termed \cite{rev3} a balancing term,
designated by ${\cal{R}}$, with the aid of which a Hilbert space operators/states that are
interrelated via the geometry have their interrelation inverted. We summarize this topic in
Section III. Thus the two succeeding sections, Section II and III, aim is to provide the
background and may be viewed
as reference for both the mathematics and notations.\\
Section IV contains the explicit expressions for the MES that play the dominant role in the
analysis. We indicate how product states that relate to points in the geometry build up
these MES that relate to (i.e. are "underpinned" by) geometrical lines. In the last
paragraphs of the section  we use Fivel's \cite{fivel} results to relate these MES with our
MUCB states, \cite{rev1}. It will be seen that, in terms of  MUCB, the proof that the $d^2$
"line" underpinned MES form an orthonormal basis for the Hilbert space is trivial. It is the
orthogonality and completeness of the MES that allows the definition of the
measurement operator in both the standard Mean King Problem (MKP)  and the King's Tracking one.\\
The derivation of our central result is given in Section VI. The first part contains the
basic and intuitively obvious result that the {\it two} particles product state
(underpinned by a geometrical point) has a non vanishing overlap with the MES (underpinned
by a line) only when the point is on the line. We then discuss the overlap in the case of a
{\it single} particle case which bears directly on the King measurement. The solution of
the MKP is now outlined and the maximally entangled state prepared by Alice in this case is
identified with the relevant balancing term (${\cal{R}}$) rather than a line underpinned
MES. Next it is proved that with the line underpinned MES as the state prepared by Alice,
the basis used by the King's  in his measurement is tracked: her control measurement allows
her to infer the basis (b), indicating that the procedure provides a novel quantum
communication channel
where the signal, sent by the King, is the identity of the basis, b, that he chose for his measurement.\\

Following Weyl \cite{weyl} and Schwinger \cite{schwinger} we use unitary operators to represent
physical quantities \cite{durt}.\\

\section{Brief Review: Mutually Unbiased Bases (MUB) and Mutually Unbiased Collective Bases (MUCB)}

In a d-dimensional Hilbert space two complete, orthonormal vectorial bases, ${\cal
B}_1,\;{\cal B}_2$,
 are said to be MUB if and only if (${\cal B}_1\ne {\cal B}_2)$
\begin{equation}
\forall |u\rangle,\;|v \rangle\; \epsilon \;{\cal B}_1,\;{\cal B}_2 \;resp.,\;\;|\langle
u|v\rangle|=1/\sqrt{d}.
\end{equation}
Maximal number of MUB allowed in a d-dimensional Hilbert space is d+1 \cite{ivanovich,bengtsson}. Variety
of methods for construction of the d+1 bases for $d=p^m$ are now available
\cite{wootters2,tal,klimov2,vourdas}. Our present study is confined to $d=p\;\ne 2$.\\
It is convenient \cite{wootters2,tal,klimov2,vourdas} to list the d+1 MUB bases in terms of the so
called computational basis (CB). The CB states $|n\rangle,\;n=0,1,..d-1,\;|n+d\rangle=|n\rangle,$
 are eigenfunction of $\hat{Z}$,
\begin{equation}
\hat{Z}|n\rangle=\omega^{n}|n\rangle;\;\omega=e^{i2\pi/d},
\end{equation}
We now give explicitly the MUB states in conjunction with the algebraically complete
operators \cite{schwinger,amir} set, $\hat{Z},$ and the shift operator,
$\hat{X}|n\rangle=|n+1\rangle:$ In addition to the CB the d other bases, each labeled by
b, are \cite{tal}
\begin{equation} \label{mxel}
|m;b\rangle=\frac{1}{\sqrt
d}\sum_0^{d-1}\omega^{\frac{b}{2}n(n-1)-nm}|n\rangle;\;\;b,m=0,1,..d-1,
\end{equation}
the m labels states within a basis. Each basis relates to a unitary operator,
\cite{tal}, $\hat{X}\hat{Z}^b|m;b\rangle=\omega^m|m;b\rangle.$ For later reference we shall
refer to the computational basis (CB) by $b=\ddot{0}$. Thus the  d+1 bases, are,
$b=\ddot{0}$ and b=0,1,...d-1. The total number of states is d(d+1) they are grouped in d+1
sets each of d states. When no confusion may arise we abbreviate the states in the CB
$|m,\ddot{0}\rangle,$ i.e. the state m in the basis $\ddot{0}$, by $|\ddot{m}\rangle,$ or
simply $|m\rangle;$ we abbreviate
$|m,0\rangle$, i.e. the m state in the basis b=0 by $|m_0\rangle$.\\
We choose the phase of the CB nil, and note  that the MUB  set is closed under complex conjugation,
\begin{eqnarray}\label{cc}
\langle n|m,b\rangle^{\ast}&=&\langle n|\tilde{m},\tilde{b}\rangle,\;\Rightarrow|\tilde{m},\tilde{b}\rangle
=|d-m,d-b\rangle,\;b\ne\ddot{0},\nonumber \\
\langle n|m\rangle&=&\langle n|m\rangle^{\ast},\;b=\ddot{0}.
\end{eqnarray}
as can be verified from Eq.(\ref{mxel}).\\
Several studies \cite{durt,klimov1,fivel,berge,rev1} consider the entanglement of two
d-dimensional particles Hilbert space via MUB state labeling. We shall now outline briefly
the approach adopted by \cite{rev1} that will be used in later sections.\\
The Hilbert space is spanned by the single particle computational bases,
$|n\rangle_1|n'\rangle_2$ (the subscripts denote the particles). These are eigenfunctions
of $\hat{Z}_i$ i=1,2:
$\hat{Z}_i|n\rangle_i=\omega^{n}|n\rangle_i,\;\omega=e^{i\frac{2\pi}{d}}.$ Similarly
$\hat{X}_i|n\rangle_i=|n+1\rangle_i,\;i=1,2$. We now define our collective coordinates and
collective operators (we remind the reader that the exponents are modular variables, e.g.
1/2 mod[d=7]=(d+1)/2)=4):
\begin{equation}\label{colz}
\hat{Z}_r\equiv \hat{Z}^{1/2}_{1}\hat{Z}^{-1/2}_{2};\;\;\bar{Z}_c\equiv
\hat{Z}^{1/2}_{1}\hat{Z}^{1/2}_{2}\;\leftrightarrow\;\hat{Z}_1=\hat{Z}_r\hat{Z}_c;\;\;\hat{Z}_2=
\hat{Z}_r^{-1}\hat{Z}_c,
\end{equation}
and, in a similar manner,
\begin{equation}\label{colx}
\hat{X}_r\equiv\hat{X}_1\hat{X}_2^{-1};\;\hat{X}_c\equiv\hat{X}_1\hat{X}_2\rightarrow
\hat{X}_1=\hat{X}^{1/2}_r\hat{X}^{1/2}_c,\;\hat{X}_2=\hat{X}^{-1/2}_r\hat{X}^{1/2}_c.
\end{equation}

Since $\bar{Z}_{s}^{d}=\bar{X}_s^{d}=1,$  and   $\bar{X}_s\bar{Z}_s=\omega\bar{Z}_s\bar{X}_s,\;s=r,c;
\;\bar{X}_s\bar{Z}_{s'}=\bar{Z}_{s'}\bar{X}_s,\;s\ne
s',$ we may consider their respective computational
eigen-bases and with it the whole set of MUB bases, \cite{rev1},
\begin{equation}
\bar{Z}_{s}|n\rangle_s = \omega^{n}|n\rangle_s,\;\; \bar{X}_s\bar{Z}_s^{b_s}|m_s,b_s\rangle=
\omega^{m_s}|m_s,b_s\rangle;\;\;\langle n_s|m_s,b_s\rangle=\omega^{\frac{b_s}{2}n_s(n_s-1)-m_sn_s}.\;s=r,c,
\end{equation}
clearly  $|n\rangle_r|n'\rangle_c;\;n,n'=0,1,..d-1,$ is a $d^2$ orthonormal basis spanning
the two d-dimensional particles Hilbert space.
One readily proves \cite{rev1}
\begin{equation}\label{relcomdel}
\langle n_1,n_2|n_r,n_c\rangle=\delta_{n_r,(n_1-n_2)/2}\delta_{n_c,(n_1+n_2)/2}.
\end{equation}
We have then,
\begin{equation}\label{relcom}
|n_r,n_c\rangle=|n_1,n_2\rangle,\;\;for\;n_r=(n_1-n_2)/2,\;n_c=(n_1+n_2)/2\;\rightleftarrows
n_1=n_r+n_c,\;n_2=n_c-n_r.
\end{equation}
There are, of course, d+1 MUB bases for each of the collective modes. Here too, we adopt the notational
simplification $b_s\rightarrow \ddot{0}_s,\; s=r,c$.

\section{   Finite Geometry and Hilbert Space Operators, Brief Outline}

The association (underpinning) scheme of finite dimensional Hilbert Space operators with
finite plane geometry that we adopt is given in \cite{rev3}. In \cite{rev3}, we studied the
underpinning of an Hilbert space single particle projectors onto an MUB state,
$$\hat{A}_{\alpha=(m,b)} \equiv |m,b\rangle\langle b,m|
\;\;m=0,1,2...d-1,\;b=\ddot{0},0,1,...d-1$$ with (the available) geometrical points,
$S_{\alpha},\; \alpha=1,2,...d(d+1):$ $S_{\alpha}\rightarrow \hat{A}_{\alpha}$. In the
present work we shall associate two particle state with a geometrical point,
\begin{equation}\label{pt}
S_{\alpha}\rightarrow |A_{\alpha=(m,b)}\rangle \equiv |m,b\rangle_1|\tilde{m},\tilde{b}\rangle_2;
\end{equation}
$|\tilde{m},\tilde{b}\rangle$ is given in Eq.(\ref{cc}), the numerical subscripts refers to
the particles. This new association requires that we reconsider the point - line
interrelation that underpin the  interrelation among the Hilbert space's two particle state, that is
underpinned by a point,
and the states that correspond to lines. To this end, for the sake of clarity  we review briefly the
essential features of finite geometry
\cite{bennett,wootters4,shirakova,tomer,saniga,rev2,rev3}:\\
A finite plane geometry is a system possessing a finite number of points and lines. There are two kinds
of finite plane geometry, affine and projective. We confine ourselves to affine plane geometry (APG).\\
It can be shown \cite{bennett,shirakova} that for $d=p^m$ (a power of prime) APG can be
constructed (our study here is for d=p,$\ne2$.  Furthermore the existence of APG implied the existence
of its
dual geometry DAPG wherein the points and lines are interchanged. Since we, in practice, utilize
DAPG we list its properties \cite{bennett, shirakova}. We shall refer to these by DAPG(.).
a. The number of lines is $d^2$, $L_j,\;j=1,2....d^2.$ The number of points is d(d+1),
$S_{\alpha},\;{\alpha = 1,2,...d(d+1)}.$\\
b. A pair of points on a line determine a line uniquely. Two (distinct) lines share one and only
one point.\\
c. Each point is common to d lines. Each line contain d+1 points: $S_{\alpha}=\bigcap_{j\in \alpha}^d L_j;
\;\;L_j=\bigcup_{\alpha \in j}^{d+1}S_{\alpha}.$\\
d. The d(d+1) points may be grouped in sets of d points no two of a set
share a line. Such a set is designated by $\alpha' \in \{\alpha \cup M_{\alpha}\},\;
\alpha'=1,2,...d$. ($M_{\alpha}$ contain all the points not connected to $\alpha$ i.e. they are not on
a line that contain $\alpha$ - they
are not connected among themselves.) i.e. such a set contain d disjoint (among themselves)
points. There are d+1 such sets:
\begin{equation}
\bigcup_{\alpha=1}^{d(d+1)}S_{\alpha}=\bigcup_{\alpha=1}^d R_{\alpha};\;\;
R_{\alpha}=\bigcup_{\alpha'\epsilon\alpha\cup M_{\alpha}}S_{\alpha'};\;\;
R_{\alpha}\bigcap R_{\alpha'}=\varnothing,\;\alpha\ne\alpha'.
\end{equation}
e. Each point of a set of disjoint points is connected to every other point not in its
set.\\

DAPG(c) allows the transcription, which we adopt, of $S_{\alpha}$ in terms of {\it
addition} of $L_j$. This acquires a meaning upon viewing the points ($S_{\alpha}$) and the
lines ($L_j$) as designating Hilbert space entities, e.g. projectors or states, which need
to be specified. (This point is further discussed at the end of this section):
\begin{equation} \label{sum}
S_{\alpha}=\frac{1}{d}\sum_{j \in \alpha}^{d}L_j\;\;\Rightarrow\;\;\sum_{\alpha{'}\in
\alpha\cup M_{\alpha}}^d S_{\alpha{'}}=\frac{1}{d}\sum_j^{d^2} L_j.
\end{equation}
DAPG(d) via Eq.(\ref{sum}) implies
\begin{equation} \label{R}
 \sum_{\alpha{'}\in \alpha\cup M_{\alpha}}^d
S_{\alpha{'}}=\frac{1}{d}\sum_j^{d^2} L_j= \frac{1}{d+1}\sum_{\alpha}^{d(d+1)}
S_{\alpha}\equiv{\cal{R}}\;\;independent\;of\;\alpha.
\end{equation}
(Eq.(\ref{R}), reflects relation among equivalent classes within the geometry
\cite{bennett}.) Thus, consistency of the transcription requires that ${\cal{R}}$ be
"universal" (i.e. independent of $\alpha$). Eq.(\ref{R}) will be referred to as the balance
formula: ${\cal{R}}$ serving as a balancing term. Thus Eqs.(\ref{sum}),(\ref{R}) imply, as
can be verified by substitution,
\begin{equation}\label{line}
L_j\;=\;\sum_{\alpha \in j}^{d+1} S_{\alpha}\;-\sum_{\alpha{'}\in \alpha\cup M_{\alpha}}^d S_{\alpha
{'}}\;=\;\sum_{\alpha \in j}^{d+1} S_{\alpha}\;-\;{\cal{R}}.
\end{equation}

A particular arrangement of lines and points that satisfies DAPG(x), x=a,b,c,d,e is
referred to as a realization of  DAPG.\\

We now consider a generic realization of DAPG of dimensionality $d=p,\ne 2$ which is the
basis of our study \cite{rev2,rev3}. We arrange the aggregate the d(d+1) points, $\alpha$,
in a $d\cdot(d+1)$matrix like rectangular array of d rows and d+1 columns. Each column is
made of a set of d points  $R_{\alpha}=\bigcup_{\alpha'\epsilon\alpha\cup
M_{\alpha}}S_{\alpha{'}};$  DAPG(d). We label the columns by $b=\ddot{0},0,1,2,....,d-1$ and
the rows by m=0,1,2...d-1.( Note that the first column label of $b=\ddot{0}$ is for
convenience with no numerical implication.)   $\alpha=m(b)$ designate a point by its row,
m, and its column, b; when b is allowed to vary - it designate the point's row position in
every column. We label the left most column by $b=\ddot{0}$ and with increasing values of
b, the
basis label, we move to the right. Thus the right most column is b=d-1.\\
Consider a realization of DAPG via Hilbert space operators or states, let ${\cal{A}}$ stand
for the Hilbert space entity underpinned with the coordinated point, (m,b). {\it In this
scheme the "universality" of ${\cal{R}}$ means that the sum along a fixed column,
$\sum_{\alpha\in b}{\cal{A}}_{\alpha=(m,b)}$ is independent of the column, b.} In
\cite{rev3} ${\cal{A}}$ stood for a projector,
${\cal{A}}_{\alpha=(m,b)}\;\rightarrow\;\hat{A}_{\alpha}=|m,b\rangle\langle b,m|.$ In the
present work ${\cal{A}}$  signifies two particles product {\it state} stipulated above,
Eq.(\ref{pt}).
 We now assert that the d+1 points, $m_j(b), b=\ddot{0},0,1,2,...d-1,$
forming the line j  contain the two (specified) points $(m,\ddot{0})$ (denoted by
$\ddot{m}$) and $(m,0)$ (denoted $m_0$) is given by (we forfeit the subscript j - it is
implicit in the two points, $j=(\ddot{m},m_0)$)),
\begin{eqnarray} \label{m(b)}
m(b)&=&\;m_0\;+\frac{b}{2}(2\ddot{m}-1),\;\;b\ne \ddot{0},\nonumber \\
m(\ddot{0})&=&\ddot{m},\;\;b= \ddot{0}.
\end{eqnarray}
These lines obviously satisfy the geometrical requirements, e.g.: i. Since
$\ddot{m},m_0=0,1,2...d-1$, there are $d^2$ lines, DAPG(a). ii. Two points determines a
line, DAPG(b). iii. Every two lines have one common point and every line contain d+1
points, DAPG(c).\\
For the underpinning considered in \cite{rev2, rev3}, viz
$S_{\alpha}\;\rightarrow\;\hat{A}_{\alpha}$ the balancing term is
${\cal{R}}=\sum_{m=0}^{d-1}|m,b\rangle\langle b,m|=\Bbb{I}$, manifestly independent of
$\alpha$, thus consistent with the balancing requirement, Eq.(\ref{R}).
 Corresponding to
$S_{\alpha}\rightarrow \hat{A}_{\alpha}$, we have via Eq(\ref{line}) $L_j\rightarrow
\hat{P}_j$, a "line" operator: $\hat{P}_j=\sum_{\alpha \in j}\hat{A}_{\alpha}-\Bbb{I}$. In
\cite{rev2,rev3} this operator is shown to abide by
\begin{equation}\label{P}
\langle n|\hat{P}_j|n'\rangle=\delta_{n+n',2\ddot{m}}\omega^{-(n-n')m_0}\;\;\hat{P}_j^2=\Bbb{I};\;
\forall\;j\;
\;tr\hat{P}_j\hat{P}_{j'}=d\delta_{j,j'}.
\end{equation}

The proof of the validity of the balance equation,  Eq.(\ref{R}), for the present case where the
transcription is   ${\cal{A}}_{\alpha=(m,b)}\;\rightarrow\;|A_{\alpha=(m,b)}\rangle$ is given in the
next section.\\

It is possible to consider different transcription corresponding to DAPG(c), \cite{tomer}, e.g.
instead of Eq.(\ref{sum}) one may take,
$$L_j\;=\;\frac{1}{d+1}\sum_{\alpha \in j}^{d+1}S_{\alpha}.$$
This transcription, though consistent, leads to more complicated formalism.\\

\section{Geometric Underpinning of Two Particles States}

We now undertake the explicit  DAPG underpinning for {\it states} in a {\it two}
d-dimensional particles Hilbert space. The coordination scheme is as given above, the
"point" $\alpha=(m,b);$ , the "line", $j=(\ddot{m},m_0).$ The line equation is given by
Eq.(\ref{m(b)}). However the point ${\alpha}$ underpins the {\it state},
$|A_{\alpha}\rangle$,
\begin{equation}\label{2p}
|A_{\alpha=(m,b)}\rangle\;=\;|m,b\rangle_1|\tilde{m},\tilde{b}\rangle_2.
\end{equation}
The numerical subscripts refers to the particles and $|\tilde{m},\tilde{b}\rangle$ is given by
Eq.(\ref{cc}). Eqs.(\ref{sum},\ref{line}) now read,
\begin{equation}
|A_{\alpha}\rangle=\frac{1}{d}\sum_{j\in\alpha}^d|P_j\rangle\;\;\Rightarrow\;\;|P_j\rangle=
\sum_{\alpha \in j}^{d+1}|A_{\alpha}\rangle\;-\;|{\cal{R}}\rangle;\;\;|{\cal{R}}\rangle=\sum_{\alpha{'}
\in \alpha\cup M_{\alpha}}|A_{\alpha{'}}\rangle.
\end{equation}
(Note that the "line" state, $|P_j\rangle$ is not normalized.) We now utilize our choice,
Eq.(\ref{cc}), to show the universality of $|{\cal{R}}\rangle$ i.e., in this case, its
independence of the basis, b, since the sum over $\alpha'=\alpha\cup M_{\alpha}$ is a sum
for a fixed b \cite{durt,fivel}:
\begin{equation}\label{uniR}
|{\cal{R}}\rangle=\sum_{\alpha{'}\in \alpha \cup M_{\alpha}}|A_{\alpha}\rangle=
\sum_{m\in b}|m,b\rangle|\tilde{m},\tilde{b}\rangle=\sum_{m,n,n'}|n\rangle_1|n'\rangle_2
\langle n|m,b\rangle \langle n'|\tilde{m},\tilde{b}\rangle=\sum_{n}|n\rangle_1|n\rangle_2,\;\;indep.
\; of\; b.
\end{equation}
The relation among the matrix elements of the projectors, $\hat{A}_{(m,b)}=|m,b\rangle\langle b,m|$,
residing on the line, Eq.(\ref{m(b)}) given in \cite{rev2,rev3} and the two particle states
$|A_{(m,b)}\rangle=|m,b\rangle|\tilde{m},\tilde{b}\rangle$, residing on the equivalent geometrical
line, Eq.(\ref{m(b)}), are now used to obtain an explicit expression for the "line" state, $|P_j\rangle$,
\begin{eqnarray}\label{line2}
|P_{j=(\ddot{m},m_0)}\rangle\;&=&\;\frac{1}{\sqrt d}\sum_{m(b)\in j}|m,b\rangle_1|\tilde{m},\tilde{b}
\rangle_2
-|{\cal{R}}\rangle= \nonumber \\
\frac{1}{\sqrt d}\sum_{n,n'}|n\rangle_1|n'\rangle_2\big[\langle n|\sum_{m(b)\in j}\hat{A}_{(m,b)}-
\Bbb{I}|n'\rangle\big]\;&=&\;\frac{1}{\sqrt d}\sum_{n,n'}|n\rangle_1|n'\rangle_2
\delta_{n+n',2\ddot{m}}\omega^{-(n-n')m_0}.
\end{eqnarray}
We used Eqs. (\ref{cc}), (\ref{P}) and (\ref{uniR}). This formula will now be put in a more pliable
form \cite{fivel} and then expressed in terms of the collective coordinates,
\begin{eqnarray}\label{list}
|P_{j=(\ddot{m},m_0)}\rangle\;&=&\;\frac{1}{\sqrt d}\sum_{n,n'}|n\rangle_1|n'\rangle_2
\delta_{n+n',2\ddot{m}}\omega^{-(n-n')m_0}=\nonumber \\ \frac{\omega^{2\ddot{m}m_0}}
{\sqrt d}\sum_n|n\rangle_1|2\ddot{m}-n\rangle_2\omega^{-2nm_0}&=&
\frac{\omega^{2\ddot{m}m_0}}{\sqrt d}\sum_n|n\rangle_1\hat{X}^{2\ddot{m}}\hat{Z}^{2m_0}
{\cal{I}}\tau|n\rangle_2 =\nonumber \\
\frac{\omega^{2\ddot{m}m_0}}{\sqrt d}\sum_m|m,b\rangle_1\hat{X}^{2\ddot{m}}\hat{Z}^{2m_0}
{\cal{I}}|\tilde{m},\tilde{b}\rangle_2&=&
|\ddot{m}\rangle_c|2m_0\rangle_r.
\end{eqnarray}
The inversion operator ${\cal{I}}$ is defined via ${\cal{I}}|n\rangle=|-n\rangle$, and we used our
definition of $\tau$ Eq.(\ref{cc}) :$\tau|m,b\rangle=|\tilde{m},\tilde{b}\rangle;\;\tau|n\rangle=
|n\rangle.$ The last equality in Eq.(\ref{list}), follows upon noting that
\begin{equation}
|\ddot{m}\rangle_c|2m_0\rangle_r=\frac{1}{\sqrt d}|\ddot{m}\rangle_c\sum_n|n\rangle_r\omega^{-2m_0n}=
\frac{1}{\sqrt d}\sum_n|\ddot{m}+n\rangle_1|\ddot{m}-n\rangle_2\omega^{-2m_0n}.
\end{equation}
That the $d^2$ vectors $|P_{j=\ddot{m},m_0}\rangle$ form an orthonormal set is manifest in the
collective coordinates formulation.\\
The central result of our geometrical approach is the following intuitively obvious overlap relation,
\begin{eqnarray}\label{cent}
\langle A_{\alpha=(m,b)}|P_{j=(\ddot{m},m_0)}\rangle\equiv \langle m,b|_1\langle\tilde{m},\tilde{b}|_2
P_{j=\ddot{m},m_0}\rangle\;&=&\;\frac{1}{\sqrt d}\delta_{m,(m_0+\frac{b}{2}[2\ddot{m}-1])},\;
b\ne\ddot{0},\nonumber \\
\langle A_{\alpha=(n,\ddot{0})}|P_{j=(\ddot{m},m_0)}\rangle\equiv\langle n|_1\langle n|_2
P_{j=\ddot{m},m_0}\rangle\;&=&\;\frac{1}{\sqrt d}\delta_{n,\ddot{m}},\;\;b=\ddot{0}.
\end{eqnarray}
Thus the overlap of $|A_{\alpha=(m,b)}\rangle$ with $|P_j\rangle$ vanishes for $\alpha\not\in j$, i.e.
for $m\ne m_0+b/2(2\ddot{m}-1)$:  Only if the point (m,b) is on the line j the overlap is not zero.
This, as shall see shortly, is the key to the solution of the MKP \cite{berge}. This is a remarkable
attribute as it holds for the particle {\it pair} while each of its constituent particle by itself
does not abide by it, Eq.(\ref{single}). We further note that the probability of finding the state
$|A_{\alpha}\rangle$ given that the state of the system is $|P_j\rangle$ with $\alpha\in j$ is 1/d
while the number of points ($\alpha$) on the line is d+1 exposing these probabilities to be not
mutually exclusive.\\

\section{Tracking the Mean King}

The maximally entangled state, Eq.(\ref{list}), was viewed geometrically as a "line", j,  state,  as
it given by the sum of product states each underpinned by the geometrical point, $\alpha=(m,b)$,
Eq.(\ref{2p}) that form the line j given by Eq.(\ref{m(b)}).  The physical, i.e. Hilbert space,
meaning of this is expressed in cf. Eq.(\ref{cent}),
\begin{equation}
|\langle A_{\alpha}|P_j\rangle|^2=|\langle \tilde{b},\tilde{m}|_2\langle b,m|_1\ddot{m}\rangle_c
|2m_0\rangle_r|^2=\begin{cases}\frac{1}{d},\;\alpha\in j,\\ 0,\;\alpha \not\in j.
\end{cases}
\end{equation}
Thus particle {\it pair}, viz. the particle and its mate, the tilde particle, whose coordinates are
$\alpha=m,b$ do as a whole belong to the d lines that share this coordinate. However each of the
constituent particles (either 1 or 2) is {\it equally} likely to be in any of the $d^2$ lines:
\begin{equation}\label{single}
\langle b,m|_1\ddot{m}\rangle_c|2m_0\rangle_r=\frac{1}{\sqrt d}|\tilde{\bar{m}}+\tilde{\Delta},
\tilde{b}\rangle_2\omega^{-2\ddot{m}\Delta},\;
\bar{m}=m_0+\frac{b}{2}(2\ddot{m}-1);\;\;\Delta=\bar{m}-m.
\end{equation}
 Thence,
\begin{equation}
|\langle b,m|_1\ddot{m}\rangle_c|2m_0\rangle_r|^2=\frac{1}{d},\;\forall\;\ddot{m},m_0.
\end{equation}
It is this attribute that allows the tracking of the King's measurement alignment.\\

We briefly outline and discuss the solution to the
 Mean King Problem (MKP) as initiated by \cite{lev}, and was analyzed in several publications,
 e.g. as listed in \cite{durt}: The state prepared by Alice is ${\cal{R}}\rangle$, Eq.(\ref{uniR}).
 The King measures in the alignment b, of his choice,  the operator,
\begin{equation}\label{K}
\hat{K}_b=\sum_{m=0}^d|m,b\rangle\omega_m\langle b,m|
\end{equation}
and observes, say $\Omega_m$. The King's measurement projects the state ${\cal{R}}\rangle$ to the
state $|A_{\alpha=m,b}\rangle$, Eq.(\ref{2p}).\\
Now for her control measurement Alice measures the non degenerate operator,
\begin{equation}\label{A}
\hat{B}\equiv\sum_{m',m"}|\ddot{m}{'}\rangle_c|2m_0^{"}\rangle_r\Gamma_{m',m"}\langle 2m_0^{"}|_r
\langle\ddot{m}{'}|_c,
\end{equation}
obtaining, say, $\Gamma_{m',m"}$. Thence the quantity,
\begin{equation}
\langle A_{\alpha=m,b}|P_{j=\ddot{m}{'},m_0{"}}\rangle\;\ne\;0,
\end{equation}
implying, cf. Eq.(\ref{cent}),
\begin{equation}\label{m"}
m=m_0{"}+(b/2)(2\ddot{m}{'}-1).
\end{equation}
Knowing $\ddot{m}{'}$ and $m_0{"}$,
through her control measurement, upon being informed of b she infers m, the King's measurement outcome,
\cite{lev,berge}. This suggest the following geometrical view. The King's measurement, Eq.(\ref{K}), of
the state prepared by Alice, ${\cal{R}}\rangle$, leaves the system at a geometrical point corresponding
to Eq.(\ref{2p}). It, cf. DAPG(c), is shared by d lines. Alice's control measurement relate a line,
$|P_{j=\ddot{m},m_0}\rangle$, to this point, Eq.(\ref{cent}). The line's equation gives m, the vectorial
coordinate, as a function of b, the basis alignment. Thus upon being told the value of b Alice can
deduce the outcome via m(b).\\
The solution of the MKP wherein Alice can specify the outcome (m) of the King measurement upon being
informed of the basis, b, he used is {\it seemingly} paradoxical \cite{werner}: She is, seemingly, able
to specify for {\it one} state, measurement outcome of each of (the possible) d+1 {\it incompatible}
bases. However this is illusionary \cite{mermin}: The relation, Eq(\ref{m"}), that gives m once b is
known holds {\it only} for the basis, b, that was actually used by the King through the presence of
$\ddot{m}{'}$ and $m_0{"}$ which relates to that basis. This relation is the central issue in Tracking
the King problem to which  we turn now.

In our case of {\it tracking} the King - he does not inform Alice the basis he used - her control
measurement is designed to track it. To this end the state that Alice prepares is one of the line
vectors $|P_{j=\ddot{m},m_0}\rangle$, Eq.(\ref{list}). Thus she knows $\ddot{m}$ and $m_0$. The
King's measurement is as in the MKP case, Eq.(\ref{K}), and, as specified above, he observed,
$\Omega_m$. In this case the King's measurement projects the state prepared by Alice,
$|P_{j=\ddot{m},m_0}\rangle$, to (neglecting normalization) to the much richer state,
\begin{equation}
|m,b\rangle_1\langle b,m|_1\ddot{m}\rangle_c|m_0\rangle_r.
\end{equation}
Now Alice, in her control measurement, measures, much like in the MKP, the operator $\hat{B}$,
Eq.{\ref{A}) and obtains, say, like in the case above, $\Gamma_{m',m"}$, implying that
\begin{equation}
\langle 2m_0{"}|_r\langle\ddot{m}{'}|_cm,b\rangle_1\langle b,m|_1\ddot{m}\rangle_c|m_0\rangle_r\;\ne\;0.
\end{equation}
The LHS of this relation is obtained via Eq.(\ref{single}) to equal, up to a  phase,
\begin{eqnarray}
\langle 2m_0{"}|_r\langle\ddot{m}{'}|_c m,b\rangle_1\langle b,m|_1\ddot{m}\rangle_c|m_0\rangle_r
&=&\frac{1}{d}\delta_{(m_0{"}-m_0),b(\ddot{m}{'}-\ddot{m})},\;b\ne\ddot{0}.
\nonumber \\
\langle 2m_0{"}|_r\langle\ddot{m}{'}|_c \ddot{n}\rangle_1\langle \ddot{n}|_1\ddot{m}\rangle_c|m_0\rangle_r
&=&\frac{1}{d}\delta_{\ddot{m},\ddot{m}^{'}},\;b=\ddot{0}.
\end{eqnarray}
Thus,
\begin{equation}\label{b}
b=\frac{(m_0{"}-m_0)}{(\ddot{m}-\ddot{m}{'})};\;\;\ddot{m}=\ddot{m}{'}\rightarrow b=\ddot{0}.
\end{equation}
Knowing the initial state, i.e. $\ddot{m}$, $m_0$, and measuring (in the control measurement)
$\ddot{m}{'}$ and $m_0{"}$ Alice tracks the King's apparatus alignment, b. The alignment $b=\ddot{0}$,
the King use of the CB, gives $\delta_{\ddot{m},\ddot{m}{'}}$.The case wherein
{\it both} $\ddot{m}{'}=\ddot{m}$ and $m_0{"}=m_0$ is undetermined.\\
Note that Alice's control measurement of $\hat{B}$, Eq.(\ref{A}), generate the state
$|P_{j=\ddot{m}{'},m_0{"}}\rangle.$ i.e. the system is reset for a succeeding message.\\

\section{ Concluding Remarks}

Two d-dimensional particles Hilbert space is analyzed in terms of center of mass and relative coordinates.
Product of states, one in the center of mass and one in he relative coordinates
that are maximally entangled (in their single particle coordinates) are used to span the Hilbert space.
These states are associated with geometrical lines and form an orthonormal basis for the space.
The formalism proves convenient in the study of the Mean King Problem (which is briefly summarized in
the paper) and a variant thereof termed Tracking the King. This, latter, case may be viewed a novel
quantum communication channel where the message sent by King to Alice is given in terms of base, b,
he used in his measurement (rather than its outcome). Alice's (subsequent) measurement that retrieve
the message sent by the King (i.e. obtain the identity of the basis he used) projects the state back
to one of the above considered maximally entangled states and thus availing it for the King's next message.\\
The relative simplicity of the formalism and its intuition aiding geometrical facet lead us to expect
(hope) that it could be of wide use in related problems.\\

 Acknowledgments: The hospitality of the Perimeter Institute where this work was completed and discussions
 with Prof. D. Gottesman, L. Hardy and M. Mueller are gratefully acknowledged. I have benefited from
 comments by Prof. W. Unruh, W. Debaere, A. Mann and an essential correction by Dr. A. Kalev.\\

\end{document}